\documentclass[preprint,aip,apl]{revtex4-1}

\usepackage{graphicx}
\usepackage{dcolumn}
\usepackage{bm}
\usepackage{subfigure}

\begin{document}

\title{Effect of epitaxial strain on the cation distribution in spinel ferrites CoFe$_2$O$_4$ and NiFe$_2$O$_4$: a density functional theory study}

\author{Daniel Fritsch}
\email{fritschd@tcd.ie}
\author{Claude Ederer}
\affiliation{School of Physics, Trinity College, Dublin 2, Ireland}

\date{\today}

\begin{abstract}
The effect of epitaxial strain on the cation distribution in spinel ferrites
CoFe$_2$O$_4$ and NiFe$_2$O$_4$ is investigated by GGA+$U$ total energy
calculations. We obtain a very strong (moderate) tendency for cation inversion
in NiFe$_2$O$_4$ (CoFe$_2$O$_4$), in agreement with experimental bulk
studies. This preference for the inverse spinel structure is reduced by
tensile epitaxial strain, which can lead to strong sensitivity of the cation
distribution on specific growth conditions in thin films. Furthermore, we
obtain significant energy differences between different cation arrangements
with the same degree of inversion, providing further evidence for recently
proposed short range $B$ site order in NiFe$_2$O$_4$.
\end{abstract}

\maketitle

The spinel ferrites CoFe$_2$O$_4$ (CFO) and NiFe$_2$O$_4$ (NFO) are
insulating ferrimagnets with high magnetic ordering temperatures and
large saturation magnetizations.~\cite{Brabers,Suzuki:2001} This
combination of properties is very attractive for a number of
applications, such as magneto-electric heterostructures and
spin-filter
devices.~\cite{Zheng_Science303_661,Dix_et_al:2010,Chapline/Wang:2006,Lueders_et_al_APL:2006,Ramos_APL91_122107}
These applications require the growth of high quality thin films of
CFO and NFO on suitable substrates. However, the electronic and
magnetic properties of the corresponding films can depend strongly on
substrate, film thickness, and specific preparation conditions, and
eventually differ drastically from the corresponding bulk
materials. For example, both increased and decreased saturation
magnetizations have been reported for thin films of CFO and NFO grown
on different substrates at different growth
temperatures.~\cite{Lueders_PRB71_134419,Rigato_MatSciEngB144_43,Ma_JAP108_063917}
It has been suggested that the large increase in magnetization
observed in some NFO films is due to the presence of Ni$^{2+}$ on the
tetrahedrally coordinated cation sites of the spinel crystal
structure.~\cite{Lueders_PRB71_134419,Rigato_MatSciEngB144_43}

The spinel crystal structure (space group $Fd\bar{3}m$) contains two
inequivalent cation sites, the tetrahedrally-coordinated $A$ sites
($T_d$) and the octahedrally coordinated $B$ sites ($O_h$). In the
\emph{normal} spinel structure, $A$ and $B$ sites are both occupied
by a unique cation species. In the \emph{inverse} spinel structure,
the more abundant cation species (Fe$^{3+}$ in the present case)
occupies the tetrahedral $A$ sites and 50\,\% of the octahedral
$B$ sites, whereas the remaining 50\,\% of $B$ sites are occupied by
the other cation species (Co$^{2+}$ or Ni$^{2+}$ in the present
case). In practice, site occupancies can vary between these two cases,
depending on specific preparation conditions, and the inversion
parameter $\lambda$ measures the fraction of less abundant cations on
the $B$ site sublattice, i.e. $\lambda=0$ for the normal spinel
structure and $\lambda=1$ for complete inversion. Since in the
ferrimagnetic N\'eel state of CFO and NFO the magnetic moments of the
$A$ and $B$ sublattices are oriented antiparallel to each other, small
changes in $\lambda$ can lead to significant changes in magnetization.

Here, we use first principles density functional theory to clarify
whether epitaxial strain can influence the distribution of cations
over the two different cation sites in CFO and NFO. Such
epitaxial strain is generally incorporated in thin films due to the
mismatch of lattice constants between the film material and the
substrate, and often leads to drastic changes of properties
compared to the corresponding bulk materials.~\cite{Suzuki:2001}

In order to accommodate different arrangements of cations on the
tetrahedral and octahedral sites in our calculations, corresponding to
different degrees of inversion, we use a unit cell described by
body centered tetragonal lattice vectors containing four formula units
(f.u.) of CFO/NFO. The unstrained cubic case corresponds to
$c/a=\sqrt{2}$. We are considering configurations corresponding to
$\lambda= \{0, 0.5, 0.75, 1\}$, and in each case (except for the
unique case $\lambda=0$) we compare at least two different
inequivalent cation arrangements. Similar to our previous
investigation we fix the internal coordinates of the cations to their
ideal values within the cubic spinel structure and fully relax the
remaining internal anion
parameters.~\cite{Fritsch_PRB82_104117,Fritsch_JPhysConfSer292_012104}
We then introduce epitaxial strain by constraining the ``in-plane''
lattice constant $a$, and relax the ``out-of-plane'' lattice constant
$c$ and all internal anion parameters. We apply strains ranging from
$-$4\,\% to $+$4\,\% relative to the relaxed $a$ lattice constant.
All our calculations are performed using the projector-augmented wave
method~\cite{Bloechl_PRB50_17953} implemented in the Vienna \textit{ab
  initio} simulation package (VASP).~\cite{Kresse_CompMatSci6_15}
We employ the generalized gradient approximation (GGA) according to
Perdew, Burke and Ernzerhof~\cite{Perdew_PRL77_3865} together with the
Hubbard ``$+U$'' correction according to Dudarev \textit{et
  al.},~\cite{Dudarev_PRB57_1505} and $U_{\text{eff}}=3$\,eV applied
to the $d$ states on all transition metal cations.
\begin{figure}[t]
\includegraphics[width=\columnwidth,clip]{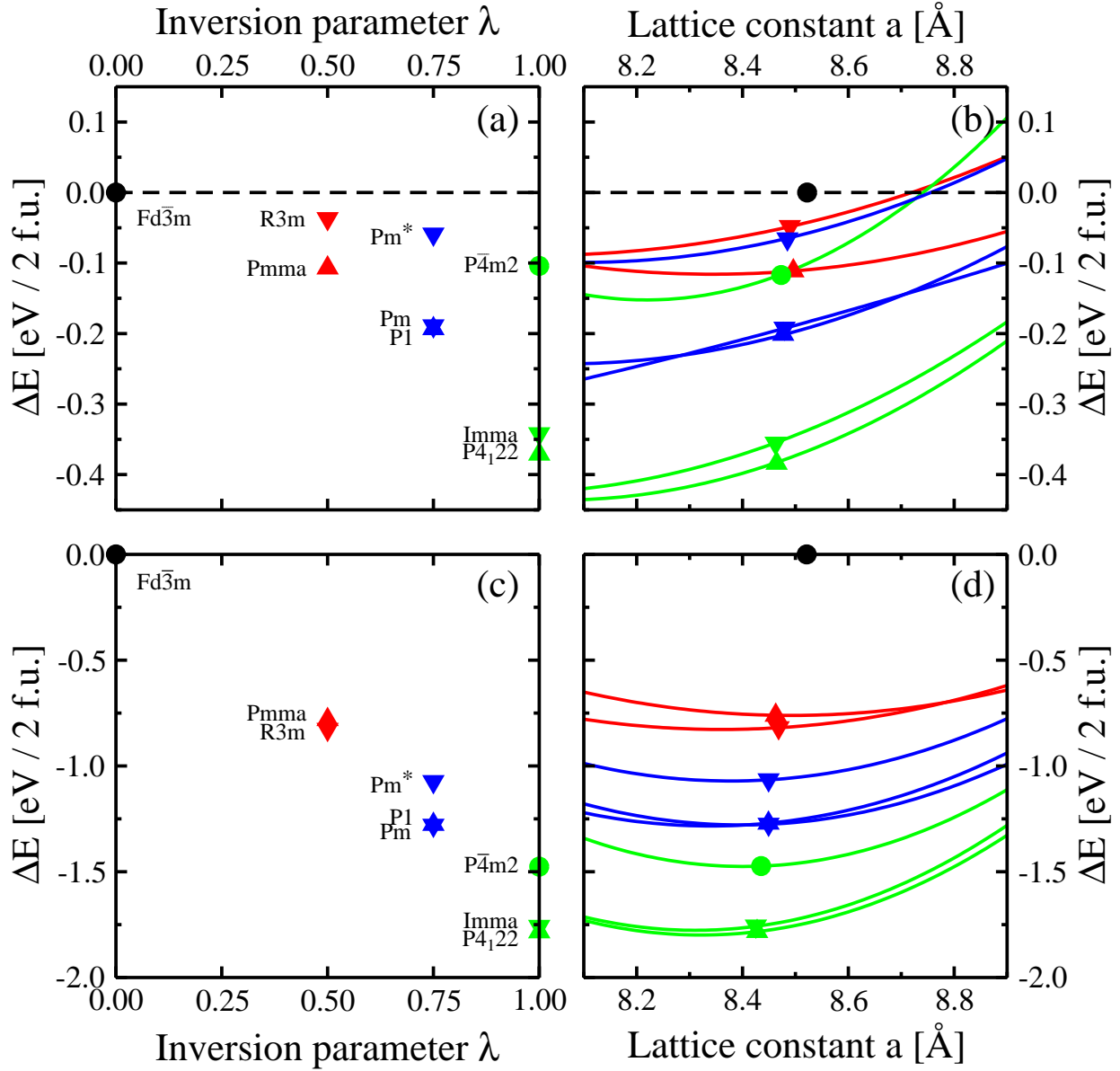}
\caption{\label{FigCFONFO_CE}(Color online) Calculated energy
  differences $\Delta E$ relative to the normal spinel structure for
  different cation arrangements corresponding to different inversion
  parameters $\lambda$ for CFO (a) and NFO (c). Variation of $\Delta E$
  with in-plane lattice constant $a$ for CFO (b) and NFO (d) in the
  various configurations. The symbols in the right panels label
  different configurations and mark the corresponding equilibrium
  lattice constant.}
\end{figure}

The calculated energy differences with respect to the normal spinel
structure for the unstrained case are shown in
Figs.~\ref{FigCFONFO_CE}(a) and \ref{FigCFONFO_CE}(c) for CFO and NFO,
respectively. Different cation arrangements are denoted by their
corresponding space group symmetry. For $\lambda=0.75$ the resulting
symmetries are very low and thus $Pm$/$Pm^*$ mark two inequivalent
configurations with the same space group. It can be seen that for both
CFO and NFO the total energy decreases with increasing inversion, so
that the fully inverse spinel structure ($\lambda=1$) is energetically
most favorable. The calculated energy difference between the normal
spinel structure and the most favorable inverse configuration is
0.37~eV (1.78~eV) per two f.u. for CFO (NFO), in good agreement with
the value of 0.339~eV reported for CFO by Hou \textit{et
  al.}~\cite{Hou_JPhysD43_445003} The much larger preference for the
inverse spinel structure of NFO compared to CFO is consistent with the
experimental observation that NFO samples usually exhibit complete
inversion, whereas the exact degree of inversion in CFO depends on the
heat treatment during sample preparation and can vary between
0.76-0.93.~\cite{Brabers,Moyer_PRB83_035121} The same energetic
preference also follows from a simple ligand-field analysis of the
Ni$^{2+}$ and Co$^{2+}$ cations within octahedral and tetrahedral
coordination.~\cite{McClure:1957,Dunitz/Orgel:1957} However, it can be
seen from our first principles results that there are also significant
energy differences between different cation arrangements corresponding
to the same value of $\lambda$. This indicates the importance of other
factors such as higher order ligand-field effects and local structural
relaxations. We note that configurations in which the Co (Ni)
cations are clustered together,
i.e. configuration $Pm^{*}$ for $\lambda=0.75$ and $P\bar{4}m2$ for
$\lambda=1$, are energetically less favorable than configurations
where Co (Ni) cations are distributed more uniformly, in agreement
with similar findings of Hou \textit{et al.}~\cite{Hou_JPhysD43_445003}

Next we turn to the question of whether the cation distribution and
degree of inversion can be influenced by epitaxial
strain. Figs.~\ref{FigCFONFO_CE}(b) and \ref{FigCFONFO_CE}(d) show the
energy differences of the strained structures relative to the energy
of the strained normal spinel at the same in-plane lattice constant
for CFO and NFO, respectively. It can be seen that the equilibrium lattice constants decrease
slightly with increasing $\lambda$. In addition, while full inversion
is most favorable for all in-plane lattice parameters, the energy
differences between different configurations decrease for larger
in-plane lattice constants. For the case of NFO the energy difference
between normal and inverse configuration reaches a maximum for $a
\approx 8.3$\,\AA\ and then decreases again for smaller in-plane
lattice constants. From these results one can expect that CFO and NFO thin films under
tensile strain are more likely to exhibit reduced inversion compared
to unstrained or compressively strained films (assuming that they are
otherwise grown under similar conditions). However, it is unclear
whether the calculated moderate changes in the relative energies will
indeed have a noticeable effect, or whether the actual cation
distribution in thin films is rather dominated by kinetic effects
related to specific growth conditions.

We also note that recent Raman investigations of both NFO single
crystals~\cite{Ivanov_PRB82_024104} and NFO thin
films~\cite{Iliev_PRB83_014108} have provided evidence for short range
cation order on the $B$ sites compatible with $P4_{1}22$ symmetry (or
equivalently $P4_{3}22$). Indeed, we find this to be the lowest energy
configuration for both CFO and NFO over the whole investigated strain
region. The energy difference compared to the $Imma$ configuration in
the unstrained structures is 28~meV (26~meV) per two f.u. for CFO
(NFO). The structural preference for $P4_{1}22$ symmetry is slightly
increased by tensile epitaxial strain in the case of NFO, whereas for
CFO the corresponding energy difference is rather independent of
strain.
\begin{figure}[t]
\includegraphics[width=0.6\columnwidth,clip]{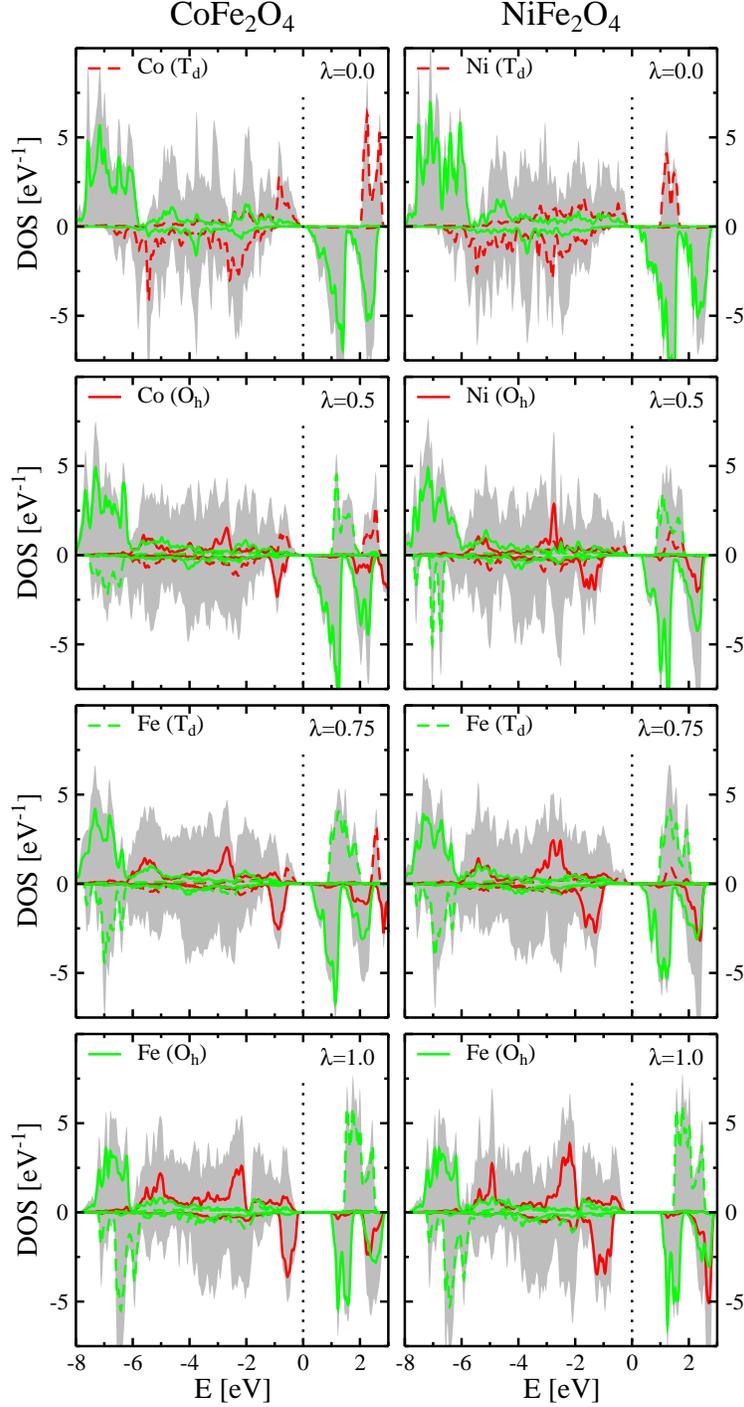}
\caption{\label{FigCFONFO_DOS}(Color online) Total and projected DOS
  per formula unit for CFO (left panels) and NFO (right panels) for
  the lowest energy configurations corresponding to different
  $\lambda$. The $d$ states of (Co, Ni) ($O_h$) and ($T_d$) are shown
  as solid and dashed red (black) lines, whereas $d$ states of Fe
  ($O_h$) and ($T_d$) are shown as solid and dashed green (dark gray)
  lines, respectively. The total DOS is shown as shaded gray area in
  all panels. Majority (minority-) spin projections correspond to
  positive (negative) values.}
\end{figure}

We now investigate the influence of cation inversion on the electronic
structure of CFO and NFO. We obtain insulating ground states for all
considered configurations. However, there is a strong tendency of the
GGA+$U$ calculations to converge to higher energy states with low-spin and/or conducting character, depending on the initial
positions for the structural relaxation.~\cite{Fritsch_InPrep}
Fig.~\ref{FigCFONFO_DOS} shows the densities of states (DOS) for CFO
and NFO corresponding to the lowest-energy configurations for each
$\lambda$. No significant differences in electronic structure for
different cation arrangements with the same $\lambda$ have been
observed. The gradual exchange of Co$^{2+}$ (Ni$^{2+}$) cations from
the $A$ site with Fe$^{3+}$ cations from the $B$ site with increasing
inversion, is reflected in the DOS by a decreasing intensity of Co
(Ni) T$_d$ and Fe O$_h$ peaks, and a corresponding increasing
intensity of Co (Ni) O$_h$ and Fe T$_d$ peaks. In addition, the band
gap of CFO (NFO) increases from 0.22~eV (0.35~eV) for $\lambda=0$ to
1.24~eV (1.26~eV) for $\lambda=1$.

The spin-splitting of the conduction band minimum (CBM), which is
important for the spin filter efficiency of magnetic tunnel junctions
containing CFO or NFO as active barrier materials, is 0.47~eV for both CFO and NFO in the fully inverse structure. This is significantly
smaller than the value of 1.28~eV (1.21~eV) for CFO (NFO), reported by
Szotek \textit{et al.},~\cite{Szotek_PRB74_174431} and is in good
agreement to recent experimental estimates in the tens of meV for
CFO-containing junctions.~\cite{Ramos_APL91_122107} The CBM
spin-splitting in CFO increases to 0.66~eV for
$\lambda=0.75$, which is closer to the value of $\lambda \sim 0.8$
observed recently in thin CFO films.~\cite{Moyer_PRB83_035121}

In summary we have analyzed the effect of epitaxial strain on the
cation distribution in the spinel ferrites CFO and NFO using first
principles total energy calculations. Using the GGA+$U$ approach we
obtain insulating electronic ground states for all degrees of
inversion and cation arrangements, and for all considered values of
epitaxial strain. We find a strong preference for the fully inverse
structure in NFO, and a somewhat weaker tendency towards cation
inversion in CFO, consistent with experimental observations. Tensile
epitaxial strain reduces this preference somewhat, which can lead to a
stronger sensitivity of the cation distribution on growth
conditions. Furthermore, for both NFO and CFO we find the (fully
inverse) $B$ site ordered arrangement with $P4_{1}22$ symmetry to be
energetically most favorable, consistent with recent experimental
results for NFO.~\cite{Ivanov_PRB82_024104,Iliev_PRB83_014108} Our
results provide a reference for the interpretation of experimental
data on CFO and NFO thin films, and thus contribute to a better
understanding of these materials as part of magnetic tunnelling
junctions and spin-filter devices.

This work was supported by Science Foundation Ireland under Ref.~SFI-07/YI2/I1051 and made use of computational facilities provided by the Trinity Centre for High Performance Computing (TCHPC) and the Irish Centre for High-End Computing (ICHEC).

%


\end{document}